\documentclass[twoside]{article}  
\usepackage{amsfonts}
\usepackage{amsbsy}
\usepackage{amsmath}
\usepackage{algorithmic}
\usepackage{algorithm}
\usepackage[pdftex]{graphicx}
\usepackage{subfig}
 \newtheorem{definition}{Definition}[section]

\newtheorem{thm}{Theorem}[section]
\newtheorem{lem}[thm]{Lemma}
\newtheorem{cor}[thm]{Corollary}
\newtheorem{example}[thm]{Example}

\newtheorem{rem}[thm]{Remark}

\newtheorem{Alg}[thm]{Algorithm}
\newcommand{\pf}{\paragraph{Proof}}
\newcommand{\pfend}{\par\vspace{2ex}\noindent}

\newcommand{\ruimte}{\par\vspace{1ex}\noindent}
\newcommand{\st}{ \;|\;}
\newcommand{\sfour}{ \;\;\;\; }

\newcommand{\Zpr}{{\mathbb Z}_{p^r}}
\newcommand{\Z}{\mathbb{Z}}
\newcommand{\F}{\mathbb{F}}
\newcommand{\dubbel}[1]{{\mathbb #1}}

\newcommand{\A}{{\cal A}}
\newcommand{\N}{\mathbb{N}}

\newcommand{\RR}{\mathcal{R}}
\newcommand{\beq}{\begin{equation}}
\newcommand{\eeq}{\end{equation}}
\newcommand{\bmat}{\left[ \begin{array}}
\newcommand{\emat}{\end{array} \right]}
\newcommand{\twee}[2]{\left[ #1 \sfour #2 \right]}
\newcommand{\COL}{\mathrm{col }\;}
\newcommand{\AND}{\;\mbox{and }}
\newcommand{\FOR}{\;\;\mbox{for }}
\newcommand{\WITH}{\;\mbox{with }}
\newcommand{\DET}{\mathrm{det }\;}

\newcommand{\DEG}{\mathrm{deg}\;}

\newcommand{\MIN}{\;\mbox{min }}
\newcommand{\ORD}{\mathrm{ord }\;}

\newcommand{\DIM}{\mathrm{dim }\;}

\newcommand{\PDIM}{p\mathrm{dim }\;}

\newcommand{\PSPAN}{p\mathrm{-span }\;}
\newcommand{\SPAN}{\;\mathrm{span }\;}

\newcommand{\Sseq}{S_1 , \ldots , S_N}
\newcommand{\ebold}{\mathbf{e}}
\newcommand{\bbold}{\mathbf{b}}

\DeclareMathOperator{\e}{e}

\DeclareMathOperator{\lt}{lt}
\DeclareMathOperator{\lc}{lc}

\DeclareMathOperator{\ord}{ord}

\DeclareMathOperator{\lm}{lm}
\DeclareMathOperator{\lpos}{lpos}
\DeclareMathOperator{\LL}{L}

\title{\LARGE \bf
An iterative algorithm for parametrization of shortest length shift registers over finite rings}
\author{M.\ Kuijper and R.\ Pinto\footnote{M.\ Kuijper is with the Department of Electrical and Electronic Engineering, University of Melbourne, VIC 3010, Australia {\tt\small mkuijper@unimelb.edu.au}; R.\ Pinto is with the Department of Mathematics, University of Aveiro, Aveiro, Portugal {\tt\small raquel@ua.pt}}
\thanks{Partly supported by the Australian Research Council(ARC); partly supported by \textit{FEDER} founds through \textit{COMPETE}--Operational Programme Factors of Competitiveness (``Programa Operacional Factores de Competitividade'') and by Portuguese founds through the \textit{Center for Research and Development in Mathematics and Applications} (University of Aveiro) and the Portuguese Foundation for Science and Technology (``FCT--Funda\c{c}\~{a}o para a Ci\^{e}ncia e a Tecnologia''), within project PEst-C/MAT/UI4106/2011 with COMPETE number FCOMP-01-0124-FEDER-022690.}}

\begin{document}

\maketitle
\begin{abstract}
The construction of shortest feedback shift registers for a finite sequence 
$S_1, \ldots , S_N$ is considered over the finite ring $\Zpr$. A novel algorithm is presented that yields a parametrization of all shortest feedback shift registers for the sequence of numbers $S_1, \ldots , S_N$, thus solving an open problem in the literature. The algorithm iteratively processes each number, starting with $S_1$, and constructs at each step a particular type of minimal Gr\"obner basis. The construction involves a simple update rule at each step which leads to computational efficiency. It is shown that the algorithm simultaneously computes a similar parametrization for the reciprocal sequence $S_N, \ldots , S_1$.
\end{abstract}

\section{Introduction}
Minimal Gr\"obner bases have been identified in the literature~\cite{fitzpatrick95,kuijperS11} as ideal tools for various types of minimal interpolation problems. Among the most fundamental of those is the problem of constructing shortest feedback shift registers for a given sequence of numbers $S_1, \ldots , S_N$. This problem is motivated by coding applications as well as cryptographic applications. In recent coding theoretic papers~\cite{wu08, kuijperA11} a parametrization of solutions is used for the purpose of list decoding of Reed-Solomon codes. In this paper we focus on the iterative construction of such a parametrization.
\ruimte
The recent paper~\cite{kuijperS11} provides a conceptual framework for a noniterative solution based on minimal Gr\"obner bases. However, the ``off the shelf'' construction of the minimal Gr\"obner basis leads to inefficiency. In this paper we aim for an iterative Gr\"obner-based solution much in the way of the efficient Berlekamp-Massey algorithm. Via a simple update rule, at each step $k$, the algorithm constructs a minimal Gr\"obner basis that yields a parametrization of all shortest feedback shift registers for the sequence $S_1, \ldots , S_k$ for $k=1,\ldots , N$. Thus the Gr\"obner basis construction is tailored to the problem at hand.
\ruimte
We find that the use of minimal Gr\"obner bases enhances the insightfulness of proofs due to the fact that we can explicitly use properties such as the 'predictable leading monomial property'', explained in the paper.
\ruimte
For the field case, the idea of a Gr\"obner based algorithm is already in the 1995 paper~\cite{fitzpatrick95}. In fact, a closer inspection shows that the algorithm of~\cite{fitzpatrick95} is practically identical to our algorithm in subsection~\ref{subsec_field} on the field case. However, our formulation differs to such an extent that it leads to a reinterpretation of some of the auxiliary polynomials as shortest feedback shift registers for the reciprocal sequence $S_N, \ldots , S_1$. This connection with the reciprocal sequence leads us to results on bidirectionality which is relevant for cryptographic applications, see~\cite{salagean09}.
\ruimte
Most importantly however, our formulation enables our main result in subsection~\ref{subsec_ring} which is an extension to sequences over the finite ring $\Zpr$ (where $p$ is a prime integer and $r$ is a positive integer). More specifically, Algorithm~\ref{alg_ring} is an iterative algorithm that constructs a parametrization of all shortest feedback shift registers for a sequence $S_1, \ldots , S_N$ in $\Zpr$. Again, the algorithm proceeds by constructing a particular type of minimal Gr\"obner basis at each step. This is where it differs from the 1985 Reeds-Sloane algorithm from~\cite{reedsS85} which also constructs a shortest feedback shift register for a sequence $S_1, \ldots , S_N$ in $\Zpr$, as a generalization of the Berlekamp-Massey algorithm. In fact, our Gr\"obner methodology yields a novel parametrization as well as insightful proofs, thus extending Massey's parametrization result to the ring case. Note that a parametrization for the $\Zpr$ case is posed as an open problem in the 1999 paper~\cite{norton99}.
\ruimte
Our proofs and results on sequences over the finite ring $\Zpr$ are nontrivial and cannot be regarded as straightforward extensions from the field case. In fact, our methodology in subsection~\ref{subsec_ring} relies heavily on a recently developed new framework~\cite{kuijperPP07,kuijperS11} for dealing with polynomial vectors in $\Zpr[x]^q$. In our earlier paper~\cite{kuijperP_it09} this methodology was applied to solve an open problem regarding minimal trellises of convolutional codes over $\Zpr$.
\ruimte
Further preliminary studies for this paper are~\cite{kuijWPifac05} and~\cite{kuijperPmtns08}.
\section{Preliminaries}
\label{sec:prelim}
Minimal Gr\"obner bases are recognized as effective tools for minimal realization and interpolation problems, see e.g.~\cite{fitzpatrick95,byrneF01,leeS08}. 
In recent papers~\cite{kuijperS11,kuijperSisit10} this effectiveness was ascribed to a powerful property of minimal Gr\"obner bases, explicitly identified as the ``Predictable Leading Monomial Property''. Before recalling this property let us first recall some terminology and basic results on Gr\"obner bases.
\ruimte
Recall that a ring is called {\em Noetherian} if
all of its ideals are finitely generated.
Let us first present some preliminaries on polynomial
vectors with coefficients in a noetherian commutative ring $\RR$. Note that $\RR [x]$ is
then also a noetherian ring.

We consider polynomials as row vectors.
Let $\e_1, \dots, \e_q$ denote the unit (row) vectors in $\RR^q$.
The elements $x^\alpha \e_i$ with $i \in \{1, \dots, q\}$ and
$\alpha \in \N_0$ are called \textbf{monomials}.
Let us consider two types of orderings on these monomials, see also the textbook~\cite{AdamsLoustaunau}:
\begin{itemize}
 \item[$\bullet$] The {\bf Term Over Position ({\tt top})} ordering, defined
 as
$$ x^{\alpha} \e_i < x^{\beta} \e_j \;\; :\Leftrightarrow \;\; \alpha < \beta
\mbox{ or } ( \alpha=\beta \mbox{ and } i<j ). $$
 \item[$\bullet$] The {\bf Position Over Term ({\tt pot})} ordering, defined
 as
$$ x^{\alpha} \e_i < x^{\beta} \e_j \;\; :\Leftrightarrow \;\; i<j \mbox{ or } ( i=j \mbox{ and }  \alpha < \beta ) .$$
\end{itemize}
Clearly, whatever ordering is chosen, every nonzero element $f \in \RR[x]^q$ can be written
uniquely as
$$ f=\sum_{i=1}^L c_i X_i, $$
where $L \in \N$, the $c_i$'s are nonzero
elements of $\RR$ for $i=1 , \ldots , L$ and $X_1, \ldots , X_L$ are
monomials, ordered as $X_1 > \dots > X_L$. Using the terminology of~\cite{AdamsLoustaunau}
we define
\begin{enumerate}
\item[$\bullet$] $\lm(f):=X_1$ as the \textbf{leading monomial} of $f$
\item[$\bullet$] $\lt(f):=c_1X_1 $ as the \textbf{leading term} of $f$
\item[$\bullet$] $\lc(f):=c_1$ as the \textbf{leading coefficient} of $f$
\end{enumerate}
Writing $X_1=x^{\alpha_1} \e_{i_1}$, where $\alpha_1 \in \N_0$ and $i_1
\in \{1,\ldots , q \}$, we define
\begin{enumerate}
 \item[$\bullet$] $\lpos(f):=i_1$ as the \textbf{leading position} of $f$
 \item[$\bullet$] $\deg(f):=\alpha_1$ as the \textbf{degree} of $f$.
\end{enumerate}
Below we denote the
submodule generated by polynomials $f_1$, ..., $f_n$ by $\langle f_1, ..., f_n \rangle$. There are several ways to define Gr\"obner bases, here we
adopt the definition of~\cite{AdamsLoustaunau} which requires us to
first define the concept of ''leading term submodule''.
\begin{definition}{\rm (\cite{AdamsLoustaunau}) }
Let $F$ be a subset of $\RR [x]^q$. Then the submodule $\LL(F)$, defined as
$$\LL(F):= \langle \lt(f) \ | \ f \in F \rangle$$
is called the \textbf{leading term submodule} of $F$.
\end{definition}
\begin{definition}\label{def_grob}{\rm (\cite{AdamsLoustaunau}) }
Let $M \subseteq \RR[x]^q$ be a module and $G \subseteq M$. Then
$G$ is called a \textbf{Gr\"obner basis} of $M$ if $$ \LL(G) = \LL(M).$$
\end{definition}
In order to define a concept of minimality we have the following
definition.
\begin{definition}\label{DefRed}{\rm (\cite[Def.\ 4.1.1]{AdamsLoustaunau}) }
Let $0 \neq f \in \RR[x]^q$ and let $F=\{f_1, \dots, f_s\}$ be a set of nonzero
elements of $\RR[x]^q$. Let $\alpha_{j_1}, \dots, \alpha_{j_m} \in
\N_0$ and $\beta_{j_1}, \dots, \beta_{j_m}$ be nonzero distinct elements of
$\RR$, where $1 \leq j_i \leq s$ for $i=1, \ldots, m$, such that
\begin{enumerate}
\item $\lm(f)=x^{\alpha_{j_i}}\lm(f_{j_i})$ for $i = 1,\ldots , m$ and
\item $\lt(f)=\beta_{j_1} x^{\alpha_{j_1}}\lt(f_{j_1}) + \dots + \beta_{j_m} x^{
\alpha_{j_m}}\lt(f_{j_m})$.
\end{enumerate}
Define
\[
h := f- (\beta_{j_1} x^{\alpha_{j_1}} f_{j_1} + \dots + \beta_{j_m} x^{\alpha_{j_m}} f_{j_m}) .
\]
Then we say that $f$ \textbf{reduces} to $h$ modulo $F$ in one step and we write

$$ f \xrightarrow{F} h . $$
If $f$ cannot be reduced modulo $F$, we say that $f$ is
\textbf{minimal} with respect to $F$.
\end{definition}
\begin{lem}\label{lemma_smaller}{\rm (\cite[Lemma 4.1.3]{AdamsLoustaunau}) }
Let $f$, $h$ and $F$ be as in the above definition. If $f
\xrightarrow{F} h$ then $h=0$ or $\lm(h)<\lm(f)$.
\end{lem}
\begin{definition}{\rm (\cite{AdamsLoustaunau})}
 A Gr\"obner basis $G$ is called \textbf{minimal} if all its elements $g$ are minimal with respect to $G\backslash\{g\}$.
\end{definition}
Elements of a minimal Gr\"obner basis have the convenient property
that all their leading monomials are different from each other. In the case that
$\RR=\F$ is a field, they have exactly $\DIM(M)$ elements and exhibit
another powerful property, see the next theorem which merely formulates a well known result.
\begin{thm}{\rm (\cite{kuijperS11})}\label{thm_plm}
Let $M$ be a submodule of $\F[x]^q$ with
minimal Gr\"obner basis $G=\{g_1 , \ldots , g_m \}$. Then for any $0 \neq f \in
M$, written as
\beq
f=a_1g_1 + \dots + a_mg_m ,\label{eq_lin_comb_f}
\eeq
where $a_1, \dots, a_m \in \F[x]$, we have
\beq
\lm(f)=\max_{1 \leq i \leq m; a_i \neq 0} ( \lm(a_i g_i)) .\label{eq_plm}
\eeq
\end{thm}
Conform~\cite{kuijperS11} we call the property of the above theorem the
\textbf{Predictable Leading Monomial (PLM) property}. Note that this property
involves not only degree information (as in the `predictable degree
property' first introduced in~\cite{forney70}) but also leading
position information. The above theorem holds no
matter which monomial ordering is chosen; here we only consider {\tt top} or {\tt pot}, but one could also employ reflected versions of {\tt top} or {\tt pot}, as in~\cite{kuijperS11} or weighted versions of {\tt top} or {\tt pot}, as 
in~\cite{kuijperA11}.
\ruimte
The next corollary follows immediately from the above theorem.
\begin{cor}\label{cor_rec_field}
Let $M$ be a submodule of $\F[x]^2$ of dimension $2$ with
minimal Gr\"obner basis $G=\{ g_1 , g_2 \}$. Suppose that $\lpos (g_2)=2$. Then $g_2$ is the lowest degree vector in $M$ with $2$ as leading position. A parametrization of all such lowest degree vectors $f$ is given by
\[
f=a_2g_2 + a_1g_1 ,
\]
with $0\neq a_2 \in \F$ and the polynomial $a_1\in \F[x]$ chosen such
that $\lm (a_1g_1) \leq \lm(g_2)$.
\end{cor}

\ruimte
Theorem~\ref{thm_plm} also leads to parametrizations of other types of minimal vectors in $M$. This is outlined in a general formulation in the next theorem.
\begin{thm}\label{thm_property}
Let $M$ be a submodule of $\F[x]^q$ with
minimal Gr\"obner basis $G=\{g_1 , \ldots , g_m \}$. Let $\ell \in \{1, \ldots ,m\}$ and let ${\cal P}$ be a property of $g_\ell$ that is absent in $\SPAN_{i\neq \ell}\{g_i\}$.
Then among all elements in $M$ with property ${\cal P}$, $g_\ell$ has minimal leading monomial. More specifically, a parametrization of all such elements is given by:
\[
f= a_\ell g_\ell + \sum_{i\neq \ell}a_i g_i ,
\]
with $0\neq a_\ell\in \F$ and for all $i\neq \ell$ the polynomials $a_i\in \F[x]$ chosen such that $\lm(a_i g_i)\leq \lm(g_\ell)$.
\end{thm}
\pf
Suppose $f \in M$ has property ${\cal P}$ and has minimal leading monomial. Obviously we can write $f$ as a linear
combination of $g_1, \ldots , g_m$. Because of the assumptions on $G$, it follows that this linear combination must use $g_\ell$. The parametrization now follows
immediately from Theorem~\ref{thm_plm}, that is, the PLM property of $G$. In particular, it follows that $\lm(f)=\lm(g_\ell)$, that is, $g_\ell$ has minimal leading monomial among all elements in $M$ with property ${\cal P}$.
\pfend
\begin{cor}\label{cor_field}
Let $M$ be a submodule of $\F[x]^2$ of dimension $2$ with
minimal Gr\"obner basis $G=\{ g_1 , g_2 \}$, where $g_1=\twee{g_{11}}{g_{12}}$ and $g_2=\twee{g_{21}}{g_{22}}$. Suppose that $g_{12}(0)=0$ and
$g_{22}(0)\neq 0$. Then $g_2$ is the lowest degree vector in $M$ that satisfies $g_{22}(0)\neq 0$. More specifically, a parametrization of all lowest degree $f=\twee{f_1}{f_2}$
in $M$ that satisfy $f_2(0)\neq 0$ is given by
\[
f=a_2g_2 + a_1g_1 ,
\]
with $0\neq a_2 \in \F$ and the polynomial $a_1\in \F[x]$ chosen such
that $\lm (a_1g_1) \leq \lm(g_2)$.
\end{cor}
\pf
Define $f=\twee{f_1}{f_2}$ to have property ${\cal P}$ if $f_2 (0) \neq 0$. The result then follows immediately from the previous theorem.
\pfend
We also have the following theorem, which merely reformulates the wellknown result of~\cite{forney75} that the maximum degree of the full size minors of a row reduced polynomial matrix equals the sum of its row degrees, see 
also~\cite{kuijperA11}.
\begin{thm}\label{thm_sum}
Let $M$ be a module in $\F[x]^q$. Let $G=\{g_1 , \ldots , g_m \}$ be a minimal Gr\"obner basis of $M$ with
respect to the {\tt top} ordering; denote the corresponding {\tt top} degrees by $\ell_i := \deg
g_i$ for $i=1, \ldots , m$. Let $\tilde G =\{\tilde g_1 , \ldots ,
\tilde g_m \}$ be a minimal Gr\"obner basis of $M$ with respect to the
{\tt pot} ordering; denote the corresponding {\tt pot} degrees by $\tilde \ell_i := \deg \tilde g_i$ for $i=1,
\ldots ,m$. Then
\beq
\sum_{i=1}^m \ell_i = \sum_{i=1}^m \tilde \ell_i .\label{eq_ell}
\eeq
\end{thm}
We call the sum in~(\ref{eq_ell}) the {\bf degree} of $M$, denoted by $\DEG (M)$.
\section{Gr\"obner bases for modules in $\Zpr[x]^q$}
\label{sec:grobner-zpr}
In this section we turn our attention to the case where $\RR$ is a finite ring of the form $\Zpr$ where $r$ is a positive integer and $p$ is a
prime integer. For the sake of completeness we repeat several
preliminaries from~\cite{kuijperPP07} and~\cite{kuijperS11}.
\subsection{Preliminaries on $\Zpr$}
A set that plays a fundamental role in this section
is the set of ``digits", denoted by $\A_p =
\{0,1,\dots, p-1\} \subset \mathbb Z_{p^r}$. Recall that any
element $a \in \dubbel{Z}_{p^r}$ can be written uniquely as
$a=\theta_0 +p \theta_1 + \cdots + p^{r-1} \theta_{r-1}$, where
$\theta_\ell \in \A_p$ for $\ell =0 , \ldots , r-1$ ({\em p-adic
expansion}).
\ruimte
Next, adopting terminology from~\cite{vaziraniSR96}, a scalar $a$ in
$\Z_{p^r}$ is said to have \textbf{order} $k$ if the additive
subgroup generated by $a$ has $p^k$ elements.
Scalars of order $r$ are called {\bf units}.
Thus the scalars $1, p, p^2, \dots, p^{r-1}$ have orders $r, r-1,
r-2, \dots, 1$, respectively. For any choice of monomial ordering ({\tt top} or {\tt pot}),
we extend the above notion of ''order'' for scalars to polynomial vectors as follows.
\begin{definition}\label{deforderpol}
The \textbf{order} of a nonzero polynomial vector $f \in \mathbb
Z_{p^r}[x]^q$, is defined as the order of the scalar $\lc (f)$, denoted as $\ORD (f)$.
\end{definition}
To deal with zero divisors occurring in $\Z_{p^r}[x]^q$,
it is useful to use notions defined in~\cite{kuijperPP07} of ''$p$-linear dependence'' and ``$p$-generator sequence'' (such notions were first introduced for "constant" modules, i.e., modules in $\Zpr^q$ in~\cite{vaziraniSR96}).

\begin{definition} {\rm (\cite{kuijperPP07})}
Let $\{v_1, \dots, v_N \} \subset \mathbb Z_{p^r}[x]^q$. A
{\bf $\boldsymbol{p}$-linear combination} of $v_1, \dots, v_N$ is a
vector $\displaystyle \sum_{j=1}^N a_j v_j,$ where $a_j
\in \A_p[x]$ for $j=1, \dots, N$. Furthermore, the set of all $p$-linear
combinations of $v_1, \dots, v_N$ is denoted by
{\bf $\boldsymbol{p}$-span}$\{v_1, \dots, v_N\}$, whereas the set of all linear
combinations of $v_1, \dots, v_N$ with coefficients in
$\mathbb Z_{p^r}[x]$ is denoted by $\SPAN \{v_1, \dots,v_N\}$.
\end{definition}

\begin{definition}\label{def_pgen} {\rm (\cite{kuijperPP07})}
An ordered sequence $(v_1, \dots, v_N)$ of vectors in
$\mathbb Z_{p^r}[x]^q$ is said to be a {\bf $\boldsymbol{p}$-generator
sequence} if $p \, v_N=0$ and $p \, v_i$ is a $p$-linear
combination of $v_{i+1}, \dots, v_N$ for $i=1, \dots, N-1$.
\end{definition}
\begin{thm}{\rm (\cite{kuijperPP07})} Let $v_1, \dots, v_N \in \mathbb Z_{p^r}[x]^q$. If
$(v_1, \dots, v_N)$ is a $p$-generator sequence then
\[
\PSPAN \{v_1, \dots, v_N\}= \SPAN \{v_1, \dots, v_N\} .
\]
In particular, $\PSPAN \{v_1, \dots, v_N\}$ is a submodule of
$\mathbb Z_{p^r}[x]^q$.
\end{thm}
All submodules of $\mathbb Z_{p^r}[x]^q$ can be written as the
$p$-span of a $p$-generator sequence. In fact, if $M= \SPAN
\{g_1, \dots, g_m\}$ then $M$ is the $p$-span of the
$p$-generator sequence
\[
(g_1,p g_1, \dots, p^{r-1}g_1,g_2,p g_2, \dots, p^{r-1}g_2,
\dots, g_m, p g_m, \dots, p^{r-1}g_m) .
\]
\begin{definition} {\rm (\cite{kuijperPP07})}
The vectors $v_1, \dots, v_N \in \mathbb Z_{p^r}[x]^q$ are
said to be {\bf $\boldsymbol{p}$-linearly independent} if the only $p$-linear
combination of $v_1, \dots, v_N$ that equals zero is the
trivial one.
\end{definition}

\begin{definition}\label{pbasis} {\rm (\cite{kuijperPP07})}
Let $M$ be a submodule of $\dubbel{Z}_{p^r}[x]^q$, written as the
  $p$-span  of a $p$-generator sequence $(v_1 , \cdots
  , v_N )$. Then  $(v_1 , \cdots , v_N )$ is
  called a
  {\bf $\boldsymbol{p}$-basis} of $M$ if the vectors $v_1, \dots,
  v_N$ are $p$-linearly independent in $\dubbel{Z}_{p^r}[x]^q$.
\end{definition}
For consistency with the field case, here we call the
  number of elements of a $p$-basis the {\bf
    $\boldsymbol{p}$-dimension} of $M$, denoted as $\PDIM(M)$.
The following definition adjusts the PLM property from the previous section to the specific
structure of $\Zpr$.
\begin{definition}{\rm (\cite{kuijperS11})}
Let $M = \PSPAN \{v_1, \dots, v_N\}$ be a submodule of $\Z_{p^r}[x]^q$. Then $\{v_1, \dots, v_N\}$ has the {\bf $\boldsymbol{p}$-Predictable
  Leading Monomial ($\boldsymbol{p}$-PLM) property} if for any $0 \neq
f \in M$, written as
\beq
f=a_1v_1 + \dots + a_N v_N ,\label{eq_plin_comb_f}
\eeq
where $a_1, \dots, a_N \in \A_p[x]$, we have
\[
\lm(f)=\max_{1 \leq i \leq N; a_i \neq 0} ( \lm(a_i f_i)) .
\]
\end{definition}
\ruimte
Note that, in contrast to the field case of the previous section, the above definition requires $a_i \in \A_p[x]$ rather than $a_i \in \RR [x]$.
\ruimte
The next theorem is the analogon of Theorem~\ref{thm_property}; we omit its proof as it is very similar to the proof of Theorem~\ref{thm_property}.
\begin{thm}\label{thm_property_ring}
Let $M = \PSPAN \{v_1, \dots, v_N\}$ be a submodule of $\Z_{p^r}[x]^q$. Assume that $\{v_1, \dots, v_N\}$ has the $p$-PLM property. Let, for some $\ell \in \{1, \ldots ,m\}$, ${\cal P}$ be a property of $v_\ell$ that is absent in $p$-linear combinations of the other $v_i$'s. Then among all elements in $M$ with property ${\cal P}$, $v_\ell$ has minimal leading monomial. More specifically, a parametrization of all such elements is given by:
\[
f= a_\ell v_\ell + \sum_{i\neq \ell}a_i v_i ,
\]
with $0\neq a_\ell\in \A_p$ and for all $i\neq \ell$ the polynomials $a_i\in \A_p [x]$ chosen such that $\lm(a_i v_i)\leq \lm(v_\ell)$.
\end{thm}
The above theorem gives rise to two corollaries. The first corollary is the ring analogon of Corollary~\ref{cor_rec_field}.
\begin{cor}\label{cor_rec_ring}
Let $M=\PSPAN\{v_1, \dots,
v_{2r}\}$ be a submodule of $\Zpr[x]^2$.
Assume that $\{v_1, \dots,
v_{2r}\}$ has the $p$-PLM property. Let $j^\star$ be such that $\lpos (v_{j^\star})=2$ and $\ord (v_{j^\star})=r$.
Then $v_{j^\star}$ is the lowest degree vector in $M$ that has order $r$ and leading position $2$. A parametrization of all such lowest degree vectors $f$ is given by
\[
f= a v_{j^\star} + \sum_{i \in \{1, \dots, 2r\} \backslash \{j^\star\}}a_i v_{i} ,
\]
with $0\neq a \in \A_p$ and for all $i \neq j^\star$ the polynomials $a_i\in \A_p[x]$ chosen such
that $\lm (a_i v_i)\leq\lm(v_{j^\star})$.
\end{cor}
\pf
Clearly all vectors in $\{v_1, \dots,v_{2r}\}$ must have either different orders or different leading position, for otherwise the $p$-PLM property would not hold. In particular, this implies that $j^\star$ is unique. Now define $f$ to have property ${\cal P}$ if $\ORD f =r$ and $\lpos (f)=2$. It follows that this property is absent in $p$-linear combinations of the $v_i$'s with $i \in \{1, \dots, 2r\} \backslash \{j^\star\}$. The result now follows from Theorem~\ref{thm_property_ring}.
\pfend
The next corollary is the ring analogon of Corollary~\ref{cor_field}.
\begin{cor}\label{cor_ring}
Let $M=\PSPAN\{v_1, \dots,
v_{2r}\}$ be a submodule of $\Zpr[x]^2$.
Assume that $\{v_1, \dots,
v_{2r}\}$ has the $p$-PLM property and write $v_i
=\twee{v_{i1}}{v_{i2}}$ for $i=1,\ldots,2r$. Also assume that
\beq
v_{i2}(0)=0 \FOR i=1,\ldots,r \AND \ORD v_{i2}(0)=2r-i+1 \FOR
i=r+1,\ldots,2r .\label{assume1}
\eeq
Then a parametrization of all lowest degree $f=\twee{f_1}{f_2}$
in $M$ with $\ORD f_2(0) =r$ is given by
\[
f=a_{r+1}v_{r+1} + \sum_{i\neq r+1}a_i v_i ,
\]
with $0\neq a_{r+1} \in \A_p$ and for all $i\neq r+1$ the polynomials $a_i\in \A_p[x]$ chosen such
that $\lm (a_i v_i)\leq\lm(v_{r+1})$.
\end{cor}
\pf
Define $f=\twee{f_1}{f_2}$ to have property ${\cal P}$ if $\ORD f_2(0) =r$, that is, $f_2 (0)$ is a unit. The result now follows immediately from Theorem~\ref{thm_property_ring}.
\pfend
The question arises whether $p$-bases with the $p$-PLM property
exist. The affirmative answer is provided by the next theorem from~\cite{kuijperS11} which is
the ring analogon of Theorem~\ref{thm_plm};
the theorem shows that the natural ordering of elements of a
minimal Gr\"obner basis gives rise to a $p$-basis with the $p$-PLM property.
\ruimte
\begin{thm}\label{thm_pGB}(\cite{kuijperS11})
Let $M$ be a submodule of $\Z_{p^r}[x]^q$ with minimal Gr\"obner basis
$G=\{g_1, \dots, g_m\}$, ordered so that $\lm(g_1) > \dots >
\lm(g_m)$. For $1 \leq j \leq m$ define
$$ \beta_j := \ORD(g_j) - \ORD(g_i),$$
where $i$ is the smallest integer $> j$ with $\lpos(g_i)=\lpos(g_j)$.
If $i$ does not exist we define $ \beta_j := \ORD(g_j)$. Then
$N=\PDIM(M)=\beta_1 + \beta_2 +\cdots + \beta_m$ and the sequence $V$ given as
\[
V=( g_1 , p g_1 , \cdots , p^{\beta_1 -1}g_1 , g_2 ,  p g_2 , \cdots ,
p^{\beta_2 -1}g_2 , \cdots,g_m, p g_m , \cdots , p^{\beta_m -1}g_m )
\]
is a $p$-basis of $M$ that has the $p$-PLM property.
\end{thm}
Conform~\cite{kuijperS11} we call $V$ a {\bf
  minimal Gr\"obner $\boldsymbol{p}$-basis} of $M$. Note that the degrees of vectors in $V$ are nonincreasing.

\section{Iterative algorithm}

Let $\RR$ be a noetherian ring, as in the previous sections. Consider a sequence $S_1, \ldots , S_N$
over $\RR$. A polynomial $\lambda (x)=\lambda_0 +\lambda_1 x + \cdots + \lambda_L x^L  \in \RR[x]$, with $\lambda_0$ a unit is called a {\bf feedback polynomial} of length $L$ if
\[
\lambda_0 S_{L + j} + \sum_{i=1}^L \lambda_i S_{L +j-i} = 0
\;\; \FOR j=1 , \ldots , N-L .\label{eq_recur}
\]
Note that $\lambda_L$ may be zero. Now consider the module $M$ in $\RR[x]^2$ defined as the rowspace of
\beq \label{matrix_M}
\bmat{c@{\hspace{2em}}c} x^{N+1} & 0 \\ -(S_N x^N + S_{N-1} x^{N-1} + \cdots + S_1 x) & 1\emat .
\eeq
We seek to find a lowest {\tt top} degree vector $\twee{\gamma (x)}{\lambda (x)}$ in $M$ for which $\lambda (0)$ is a unit. In terms of trajectories this vector can be interpreted as an annihilator: we have $\twee{-\gamma (\sigma)}{\lambda (\sigma)}\bbold =0$, where $\sigma$ is the forward shift operator and $\bbold: {\dubbel{Z}}_- \mapsto \RR^2$ is given by
\beq
\bbold := \left(\ldots \ \bmat{c} 0
\\ 0 \emat, \bmat{c} 1 \\ 0 \emat ,\bmat{c} 0 \\ S_1 \emat , \bmat{c} 0 \\ S_2 \emat, \ldots ,
\bmat{c} 0 \\ S_N \emat \right) .\label{bdef}
\eeq
Our objective in this paper is to develop an iterative algorithm to
construct feedback polynomials of shortest length. This length is
called the {\bf complexity} of the sequence. We require the algorithm
to construct, at each step $k$, an annihilator for
$\sigma^{N-k}\bbold$ of lowest {\tt top} degree.
\begin{rem}\label{remark}
{\rm Note that the requirement to process $S_1, \ldots , S_k$ at step $k$
(rather than $S_N, \ldots , S_{N-k+1}$) necessitates our formulation
in terms of ``feedback polynomial'' $\lambda$, rather than its
reciprocal version, denoted as $d$ in~\cite{kuijperS11}.
In this paper
we call $d$ a characteristic polynomial
of the sequence $S_1 ,\ldots , S_N$; the degree of a minimal characteristic polynomial equals the complexity
of the sequence. Thus, a polynomial $d$ written as $d(x)=d_L x^L+ \cdots +d_0$ is a {\bf characteristic polynomial} of
$S_1 ,\ldots , S_N$ if $d_L$ is a unit and
\[
d_L S_{L +j} + \sum_{i=1}^{L}d_{L-i}
S_{L +j-i}=0\FOR j=1,\ldots,N-L .
\]
Consider the reciprocal module $M^{rec}$ in $\RR[x]^2$, defined as the rowspace of
\beq \label{matrix_M_rec}
\bmat{c@{\hspace{2em}}c} x^{N+1} & 0 \\ -(S_1 x^N + S_2 x^{N-1} + \cdots + S_N x) & 1\emat .
\eeq
It is easily verified that a minimal characteristic polynomial $d$ for $S_1
,\ldots , S_N$ is found in any vector $\twee{h}{d}$ in $M^{rec}$ of leading position $2$ that has minimal leading monomial, see~\cite{kuijperS11}.
Note that, by definition, whenever $\lambda_L$ is a unit, a feedback polynomial $\lambda (x)=\lambda_0 +\lambda_1 x + \cdots + \lambda_L x^L $ of length $L$ for a sequence $S_1,\ldots, S_N$ also serves as a characteristic polynomial of the reciprocal sequence $S_N,\ldots, S_1$; such a polynomial is called {\bf bidirectional} as in~\cite{salagean09}, see also~\cite{norton10}.}
\end{rem}
\subsection{The field case}\label{subsec_field}
In this subsection we focus on the case that $\RR$ is a field $\F$.
Note that $\F$ is not required to be finite.
The Berlekamp-Massey algorithm is a famous iterative algorithm
that constructs a feedback polynomial of shortest length for a sequence $S_1,
\ldots , S_N$ in $\F$. It processes a new data element $S_k$ at each
step $k$ for $k=1, \ldots , N$ and then produces a feedback polynomial of shortest length for $S_1, \ldots , S_k$. In this subsection we present an algorithm that is identical to the
Berlekamp-Massey algorithm apart from a slight modification of the
update rule. Due to this modification, our algorithm iteratively
constructs a minimal Gr\"obner basis at each step. The algorithm shares several useful
properties with the Berlekamp-Massey algorithm, namely that it
processes the data in a natural order and that it allows us to read off the solution at once.
A closer inspection shows that the algorithm of~\cite{fitzpatrick95} is practically identical to our algorithm. Our formulation is different however, using $2 \times 2$ polynomial matrices, as in Berlekamp's original work~\cite{berlekamp68}, see also its formulation in the textbook~\cite{blahut83}. This formulation facilitates explicit use of the PLM property yielding a parametrization of all solutions as well as a result on the reciprocal sequence, see Theorem~\ref{main_paraBM} below. Furthermore, this formulation facilitates an extension to sequences over the finite ring $\Zpr$, presented in subsection~\ref{subsec_ring} below. 
This extension proves nontrivial as it involves a careful use of the minimal Gr\"obner $p$-bases of the previous section.
\ruimte
In this subsection we focus on modules in $\F[x]^q$, where $\F$ is a field. In section~\ref{sec:prelim} minimal Gr\"obner bases were defined, they can be computed for any module using computational packages such as {\tt Singular}. Here we will not use such packages, instead we iteratively construct minimal Gr\"obner bases in a computationally efficient way. In order to be able to do this we first need to answer the following question: given a set of vectors in $M$, how do we recognize this set as a minimal Gr\"obner basis? The next theorem considers the special case in which $M$ is a full rank module; the theorem holds for either {\tt top} or {\tt pot} monomial ordering and uses the definition of a module's ``degree'' following from Theorem~\ref{thm_sum}.
\begin{thm}\label{thm_GBchar}
Let $M \in \mathbb F[x]^q$ be a module of dimension $q$ and degree
$\delta$ and let $G=\{g_1(x), \dots, g_q(x)\} \subset M$. Then
$G$ is a minimal Gr\"obner basis of $M$ if
and only if the following two conditions hold:
\begin{enumerate}
\item $\displaystyle \sum_{i=1}^q \DEG g_i = \delta$;

\item all leading positions of the vectors $g_1(x), \dots, g_q(x)$
are different.
\end{enumerate}
\end{thm}
\pf
Let $\tilde G =\{\tilde g_1, \dots, \tilde g_q\}$ be a minimal Gr\"obner basis of $M$, ordered
as $\lm(\tilde g_1) > \lm(\tilde g_2) > \cdots > \lm(\tilde g_q)$. Let $i \in \{1, \dots,q\}$. It is obvious that all leading positions of $\tilde g_1(x), \dots, \tilde g_q(x)$ are different. Without restrictions we may therefore assume that $g_i$ and
$\tilde g_i$ have the same leading position.
The predictable leading monomial property of $\tilde G$ (see Theorem~\ref{thm_plm}) now implies that $g_i$ is a linear combination of $\tilde
g_1, \dots, \tilde g_q$ that uses $\tilde g_i$ and it follows that $\lm(g_i)
\geq \lm(\tilde g_i)$. Since $g_i$ and
$\tilde g_i$ have the same leading position, this implies that $\DEG g_i \geq \DEG \tilde
g_i$. Consequently, by condition 1) of the theorem
\[
\delta = \displaystyle \sum_{i=1}^q
\DEG g_i \geq \displaystyle \sum_{i=1}^q \DEG \tilde g_i = \delta ,
\]
so that it follows that $\DEG g_i = \DEG \tilde g_i $ for $i=1, \dots, q$. We also conclude
that $\lt(g_i) = a_i \, \lt(\tilde g_i)$ for some $0\neq a_i \in \F$ for $i=1, \dots, q$.
As a result $L(G)=L(\tilde G)=L(M)$, so that, by Definition~\ref{def_grob}, $G$ is a Gr\"obner basis for
$M$. Furthermore, clearly $G$ cannot be reduced, so that $G$ is a minimal Gr\"obner basis for $M$.
 \pfend
In the next algorithm the unit vectors $e_1$ and $e_2$ are defined as $e_1:=\twee{1}{0}$ and $e_2:=\twee{0}{1}$; the two rows of the matrix $R^k$ are denoted by $g_1^k$ and $g_2^k$, respectively. Recall that $\sigma$ denotes the forward shift operator.
\begin{Alg}\label{alg_field}
 {\bf Input data:} $\Sseq$.

{\bf Initialization:} Define
$$
R^0(x):=\bmat{cc} x & 0  \\ 0 & 1 \emat.
$$

{\bf Proceed iteratively as follows for $k=1, \dots, N$.}

\begin{itemize}
\item Define the error trajectory
\[
\ebold^k := (\ldots,0,0,\Delta^k ) := R^{k-1}(\sigma ) \bbold_k ,
\]
where $\bbold_k$ is given as $\bbold_k := \sigma^{N-k} \bbold$, with $\bbold$ given by~(\ref{bdef}).
\item Denote $\Delta^k=\bmat{cc} \Delta^k_1 & \Delta^k_2 \emat^T$.
\item Define ${\cal P}^k :=\{i \in \{1,2\}:
\Delta^k_i \neq 0\}$.
\item Define $i^\star := \mbox{arg} \, \displaystyle \MIN_{i \in
{\cal P}^k} \{\lm(g_i^{k-1})\}$.
\item Define the update matrix
$E^k(x) := \frac{x}{\Delta^k_{i^*}}e_1^T e_{i^\star} + e_2^T(-\Delta^k_2 e_1+\Delta^k_1 e_2)$.
\item Define $R^k(x):=E^k(x)R^{k-1}(x)$.

\end{itemize}

{\bf Output:} $R(x):= R^N(x)$.

\end{Alg}
\begin{lem}\label{lem_properties_field}
Let $\Sseq$ be a sequence over a field $\mathbb F$ and let $k \in \{0,\ldots , N\}$. Let $R^k$ be the matrix
obtained by applying Algorithm \ref{alg_field} to
$S_1, \ldots , S_k$. Denote the two rows of $R^k$ by $g_1^k := [g_{11}^k \;\; g_{12}^k]$ and $g_2^k := [g_{21}^k \;\; g_{22}^k]$. Then, with respect to the {\tt top} monomial ordering:
\begin{itemize}
\item[i)] $\DEG g_1^k + \DEG g_2^k = k+1$
\item[ii)] $\lpos(g_1^k ) \neq \lpos(g_2^k )$
\item[iii)] $g_{1}^k (0) =\twee{0}{0}$ and $g_{22}^k (0) =1$
\item[iv)] $\Delta_1^{k+1} =1$
\item[v)] $R^k (\sigma ) \bbold_k =0$
\end{itemize}
\end{lem}
\pf
Clearly the lemma holds for $k=0$. Let us now proceed by
induction and assume that the lemma holds for some $k \in \{0,1,\ldots, N-1\}$.
To prove (iv) we observe that, by definition, $g_1^{k+1}(x)=\frac{x}{\Delta_1^{k+1}} g_1^{k}(x)$ if $i^*=1$, or $g_1^{k+1}(x)=\frac{x}{\Delta_2^{k+1}} g_2^{k}(x)$ if $i^*=2$. Thus, if $i^*=1$ then
\[
g_1^{k+1}(\sigma) \bbold_{k+2}=\frac{1}{\Delta_1^{k+1}}g_1^{k}(\sigma) \sigma \bbold_{k+2}=\frac{1}{\Delta_1^{k+1}}g_1^{k}(\sigma) \bbold_{k+1}= \frac{1}{\Delta_1^{k+1}}(\ldots,0,0,\Delta_1^{k+1} )=(\ldots,0,0,1),
\]
in other words $\Delta_1^{k+2} =1$. Similarly, if $i^*=2$ then also $g_1^{k+1}(\sigma) \bbold_{k+2}=(\ldots,0,0,1)$, so that also in this case $\Delta_1^{k+2} =1$.
To prove (v), observe that $R^{k+1}(x):=E^{k+1}(x)R^k(x)$ so that $R^{k+1}(\sigma)\bbold_{k+1} =
E^{k+1}(\sigma)R^k(\sigma)\bbold_{k+1} =  E^{k+1}(\sigma)\ebold_{k+1}$
equals the zero trajectory by definition of $E^{k+1}$. Thus (v) holds. Further, using again the definition of $E^{k+1}$ as well as
the induction hypotheses, it follows that (i)-(iii) hold. Thus all
properties hold for $k+1$ and this proves the lemma by induction.
\pfend

\begin{thm}\label{main_paraBM}
Let $\Sseq$ be a sequence over a field $\mathbb F$ and let $R$ be the matrix
obtained by applying Algorithm \ref{alg_field} to
$S_1, \ldots , S_N$. Denote the two rows of $R$ by $g_1=[g_{11} \;\; g_{12}]$ and $g_2=[g_{21} \;\; g_{22}]$;
denote $\tilde L:= \DEG g_1$ and $L:= \DEG g_2$ with respect to the {\tt top} monomial ordering.
Then the complexity of the sequence equals $L$ and $g_{22}$ is a feedback polynomial of shortest length $L$. More specifically, a parametrization of all shortest length feedback polynomials is given by
\beq
a g_{22} + b g_{12},\label{eq_para_field}
\eeq
where $0\neq a\in \F$ and $b\in \F [x]$ such that $\DEG b \leq
L-\tilde L$.
\ruimte
Furthermore, if $\lpos(g_2)=2$ then the feedback polynomial $g_{22}$ is bidirectional and~(\ref{eq_para_field}) also parametrizes all bidirectional minimal characteristic polynomials of the reciprocal sequence $S_N, \ldots , S_1$.
Otherwise, i.e. if $\lpos(g_2)=1$ then the complexity of the reciprocal sequence
$S_N, \ldots , S_1$ equals $\tilde L$ and $g_{12}$ is a minimal characteristic polynomial of $S_N, \ldots , S_1$. More specifically, a parametrization of all minimal characteristic
polynomials of $S_N, \ldots , S_1$ is given by
\beq
a g_{12} + b g_{22},\label{eq_para_reciprocal_field}
\eeq
where $0\neq a\in \F$ and $b\in \F [x]$ such that $\DEG b \leq
\tilde L - L$. In particular, any choice of $b\in \F [x]$ such that $\DEG b \leq
\tilde L - L$ and $b(0) \neq 0$ gives a bidirectional minimal characteristic
polynomial of $S_N, \ldots , S_1$. 
\end{thm}
\pf
Let $M$ be defined as the row space of~(\ref{matrix_M}). From Theorem~\ref{thm_GBchar}
and (i) and (ii) of the previous lemma, it follows that for all $k \in
\{0,\ldots , N\}$ the set $\{ g_1^k , g_2^k \}$ is a minimal Gr\"obner basis for the row space of
\[
\bmat{c@{\hspace{2em}}c} x^{k+1} & 0 \\ -(S_k x^k + S_{k-1} x^{k-1} + \cdots + S_1 x) & 1\emat .
\]
The theorem now follows immediately from
Corollary~\ref{cor_field}. The statements on the reciprocal sequence
follow immediately from Remark~\ref{remark} and Corollary~\ref{cor_rec_field}
(note that $\lpos(g_2)=1$ implies that $\lpos(g_1)=2$).
\pfend
\ruimte
\begin{example}
Consider the sequence $S_1,S_2,S_3,S_4,S_5=4,0,4,4,2$ over the field $\Z_5$. Application of Algorithm~\ref{alg_field} yields:
$$\begin{array}{llll}
\Delta^1=\bmat{c} 1 \\ 4 \emat, & {\cal P}^1=\{1,2\}, & i^*=2, & R^1(x)= \bmat{cc} 0 & 4x \\ 1 & 1 \emat R^0(x)=\bmat{cc} 0 & 4x \\ x & 1 \emat; \\
\Delta^2=\bmat{c} 1 \\ 0 \emat, & {\cal P}^2=\{1\}, & i^*=1, & R^2(x)= \bmat{cc} x & 0 \\ 0 & 1 \emat R^1(x)=\bmat{cc} 0 & 4x^2 \\ x & 1 \emat; \\
\Delta^3=\bmat{c} 1 \\ 4 \emat, & {\cal P}^3=\{1,2\}, & i^*=2, & R^3(x)= \bmat{cc} 0 & 4x \\ 1 & 1 \emat R^2(x)=\bmat{cc} 4x^2 & 4x \\ x & 4x^2+1\emat; \\
\Delta^4=\bmat{c} 1 \\ 4 \emat, & {\cal P}^4=\{1,2\}, & i^*=1, & R^4(x)= \bmat{cc} x & 0 \\ 1 & 1 \emat R^3(x)=\bmat{cc} 4x^3 & 4x^2 \\ 4x^2+x & 4x^2+4x+1 \emat;\\
\Delta^5=\bmat{c} 1 \\ 4 \emat, & {\cal P}^5=\{1,2\}, & i^*=2, & R^5(x)= \bmat{cc} 0 & 4x \\ 1 & 1 \emat R^4(x)=\bmat{cc}  x^3+4x^2 & x^3+x^2+4x \\ 4x^3+4x^2+x & 3x^2+4x+1 \emat.
\end{array}$$
By the above theorem, the complexity of the sequence equals $L=3$ and  $3x^2+4x+1$ is a shortest length feedback polynomial.
The complexity of the reciprocal sequence $2,4,4,0,4$ equals $\tilde L=3$ and  $x^3+x^2+4x$ serves as a minimal characteristic polynomial of $2,4,4,0,4$. 
From the parametrization~(\ref{eq_para_reciprocal_field}) we see that there is only one monic bidirectional minimal characteristic polynomial with value $1$ at $x=0$, namely $(x^3+x^2+4x)+(3x^2+4x+1)=x^3+4x^2+3x+1$.
\end{example}
\begin{rem}
{\rm The earlier paper~\cite{kuijWac97} formulates the Berlekamp-Massey
algorithm in a similar format as Algorithm~\ref{alg_field}. From this
it is clear that Algorithm~\ref{alg_field} differs from the Berlekamp-Massey
algorithm only in the definition of $i^\star$. More precisely, in the Berlekamp-Massey
algorithm $i^\star$ equals the largest integer $i$ in ${\cal P}^k$ such that $g_i^{k-1}$ has minimal degree.  Application of the Berlekamp-Massey
algorithm in the above example gives the same first three steps leading to $R^3(x)= \bmat{cc} 4x^2 & 4x \\ x & 4x^2+1\emat$. However, the next two steps give a different result:
$$\begin{array}{llll}
\Delta^4=\bmat{c} 1 \\ 4 \emat, & {\cal P}^4=\{1,2\}, & i^*=2, & R^4(x)= \bmat{cc} 0 & 4x \\ 1 & 1 \emat R^3(x)=\bmat{cc} 4x^2 & x^3+4x \\ 4x^2+x & 4x^2+4x+1 \emat;\\
\Delta^5=\bmat{c} 1 \\ 4 \emat, & {\cal P}^5=\{1,2\}, & i^*=2, & R^5(x)= \bmat{cc} 0 & 4x \\ 1 & 1 \emat R^4(x)=\bmat{cc}  x^3+4x^2 & x^3+x^2+4x \\ 3x^2+x & x^3+4x^2+3x+1 \emat.
\end{array}$$
In particular we see that here the rows of $R^4$ do not constitute a minimal Gr\"obner basis, since both rows have leading position $2$. Similarly, the rows of $R^5$ do not constitute a minimal Gr\"obner basis. Thus this example illustrates a main difference between the Berlekamp-Massey
algorithm and our Algorithm~\ref{alg_field}: by keeping track of leading position information, our algorithm produces a minimal Gr\"obner basis, whereas the Berlekamp-Massey algorithm does not necessarily produce a minimal Gr\"obner basis since it only keeps track of degree information. The advantage of the Gr\"obner formulation is that it allows for a transparent extension to the ring case, as detailed in the next subsection.
}
\end{rem}

\subsection{The ring case}\label{subsec_ring}

In this subsection we present our main result which is an
algorithm that extends the algorithm from the previous subsection to
the ring case. We focus on a finite sequence $S_1, \ldots , S_N$ from
$\Zpr$ and seek to construct a feedback polynomial of shortest length
(including parametrization) by iteratively processing the data in the
natural order $S_1, \ldots , S_N$. Again our key object of interest is
the module $M$ defined as the row space of~(\ref{matrix_M}). Our algorithm
constructs a $2r \times 2$ polynomial matrix $R$ whose rows are a
$p$-basis for $M$ that has the $p$-PLM property.
\ruimte
In the next algorithm the $2r$ rows of the matrix $R^k$ are denoted by $v_1^k,\ldots,v_{2r}^k$.

\begin{Alg}\label{alg_ring}
 {\bf Input data:} $\Sseq$.

{\bf Initialization:} Define
$$
R^0(x):=\bmat{cc} x & 0  \\ \vdots & \vdots \\
 p^{r-1}x & 0 \\ 0 & 1 \\ \vdots & \vdots \\ 0 &  p^{r-1} \emat.
$$

{\bf Proceed iteratively as follows for $k=1, \dots, N$.}

\begin{itemize}
\item Define the error trajectory
\[
\ebold^k := (\ldots, 0,0,\Delta^k) := R^{k-1}(\sigma ) \bbold_k ,
\]
where $\bbold_k$ is given as $\bbold_k := \sigma^{N-k} \bbold$, with $\bbold$ given by~(\ref{bdef}).
\item Denote $\Delta^k=\bmat{ccc} \Delta^k_1 & \cdots &
\Delta^k_{2r} \emat^T$ and let $\theta^k_i$ be a unit and
$\ell^k_i \in \{1, \dots, r\}$ such that $\Delta^k_i =
\theta^k_i p^{\ell^k_i -1}$ for $i=1, 2, \dots, 2r$.
\item Define ${\cal P}^k_0:=\{i \in \{1 ,\dots, 2r\}:
\Delta^k_i=0\}$.
\item For $j=1, \dots, r$, define ${\cal P}^k_j:=\{i \in \{1,
\dots, 2r\}: \ell^k_i=j\}$.
\item Define $i^\star_j$ as the largest index $i$ in ${\cal P}^k_j$ for which $\lm(v^{k-1}_i)$ is minimal.
\item Define the update matrix $E^k(x)$ as
\[
E^k(x):=\displaystyle \sum_{i \in {\cal P}^k_0} e_i^T
e_i + \displaystyle \sum_{j=1}^r \left[
\frac{x}{\theta^k_{i^*_j}}e_j^T e_{i^\star_j} + e_{i^\star_j}^T(-\theta^k_{i^\star_j}e_j+
\theta^k_j e_{i^\star_j}) + \displaystyle \sum_{i \in {\cal P}^k_j \backslash
\{j, i^\star_j\}} e_i^T (- \theta^k_i e_{i^\star_j}+ \theta^k_{i^\star_j}
e_i)\right].
\]
\item Define $R^k(x):=E^k(x)R^{k-1}(x)$.

\end{itemize}

{\bf Output:} $R(x):= R^N(x)$.
\end{Alg}
\begin{lem}\label{lem_properties_ring}
Let $\Sseq$ be a sequence over $\Zpr$ and let $k \in \{0,\ldots , N\}$. Let $R^k$ be the matrix
obtained by applying Algorithm \ref{alg_ring} to
$S_1, \ldots , S_k$. Denote the rows of $R^k$ by $v_1^k,\ldots,v_{2r}^k$; denote $v_j^k := \twee{v_{j1}^k}{v_{j2}^k}$ for $j=1,\ldots , 2r$. Then, with respect to the {\tt top} monomial ordering:
\begin{itemize}
\item[i)] $\DEG v_1^k + \ldots + \DEG v_{2r}^k = r(k+1)$
\item[ii)] if $i, j \in \{1, \dots, 2r\}$, with $i \neq j$, then
  $\lpos(v^k_i)=\lpos(v^k_j) \Rightarrow \ord(v^k_i) \neq  \ord(v^k_j)$
\item[iii)] $\Delta_j^{k+1}=p^{j-1}$ for $j=1, \dots, r$
\item[iv)] $R^k (\sigma ) \bbold_k =0$
\end{itemize}
\end{lem}
\pf
Clearly the lemma holds for $k=0$. Let us now proceed by
induction and assume that the lemma holds for some $k \in \{0,1,\ldots
, N-1\}$.
To prove (iii), let $j \in \{ 1, \ldots , r\}$. By definition, $v^{k+1}_j(x)=\frac{x}{\theta^{k+1}_j} v^k_j(x)$ if $i^*_j=j$, and $v^{k+1}_j(x)=\frac{x}{\theta^{k+1}_{i^*_j}} v^{k}_{i^*_j}(x)$, otherwise. Thus in case $i^*_j=j$ we have
\[
v^{k+1}_j(\sigma) \bbold_{k+2}=\frac{1}{\theta^{k+1}_j}v^{k}_j(\sigma) \sigma \bbold_{k+2} = \frac{1}{\theta^{k+1}_j}v^{k}_j(\sigma) \bbold_{k+1} = (\ldots ,0, 0,p^{j-1}) .
\]
In case $i^*_j \neq j$ it follows in an entirely similar way that $v^{k+1}_j(\sigma) \bbold_{k+2}= (\ldots ,0, 0,p^{j-1})$. Thus (iii) holds.
To prove (iv), note that $R^{k+1}(\sigma)\bbold_{k+1} =
E^{k+1}(\sigma)R^k(\sigma)\bbold_{k+1} =  E^{k+1}(\sigma)\ebold_{k+1}$
equals the zero trajectory by definition of $E^{k+1}$. In other words,
(iv) holds. 
By definition, in the update operation $R^{k+1}=E^{k+1} R^k$ the degrees of exactly $r$ rows of $R^k$ are increased by $1$, so that (i) holds by induction. Similarly, it is easily seen from the definition of $E^{k+1}$ that (ii) holds by induction.
\pfend
\begin{lem}\label{lem_pPLM}
Let $\Sseq$ be a sequence over $\Zpr$, let $M$ be the module defined
as the row space of~(\ref{matrix_M}). Let $R$ be the matrix
obtained by applying Algorithm \ref{alg_ring} to
$S_1, \ldots , S_N$. Denote the rows of $R$ by $v_1, \ldots
,v_{2r}$.
Then $\{v_1 , \ldots , v_{2r}\}$ is a $p$-basis of $M$ that has the $p$-PLM property.
\end{lem}
\pf
Let $\tilde V=(\tilde v_1 , \ldots , \tilde v_{2r})$ be a minimal Gr\"obner
$p$-basis of $M$, as defined in Theorem~\ref{thm_pGB}. Note that, by definition,
\beq
\lm (\tilde v_{i+1}) \leq \lm (\tilde v_i) \FOR i=1,\ldots ,2r-1 .\label{decreasing_degrees_pgrob}
\eeq
and for $i < j$ we have
\beq
\lm(\tilde v_i)=\lm(\tilde v_j) \Rightarrow \ord(\tilde v_i) > \ord(\tilde v_j).\label{decreasing_orders_pgrob}
\eeq
As a result, defining $G_1:=\{\tilde v\in \tilde V \st \; \lpos(\tilde v) = 1\}$ and $G_2:=\{\tilde v\in \tilde V \st \; \lpos(\tilde v) = 2\}$, there exists a bijection $\phi: G_1 \rightarrow G_2$ such that $\ord (\phi(\tilde v))= r + 1 - \ord(\tilde v)$ for all $\tilde v\in G_1$.
Clearly $\DEG \DET \COL(\tilde v,\phi(\tilde v))=\DEG \tilde v +\DEG \phi(\tilde v)$ for all $\tilde v\in G_1$. On the other hand,
\[
\COL(\tilde v,\phi(\tilde v)) = U(x) \bmat{c@{\hspace{2em}}c} x^{N+1} & 0 \\ -(S_N x^N + S_{N-1} x^{N-1} + \cdots + S_1 x) & 1\emat,
\]
for some polynomial matrix $U(x)$, so that $\DEG \tilde v +\DEG \phi(\tilde v)\geq N+1$ for all $\tilde v\in G_1$. As a result,
\beq
\sum_{i=1}^{2r} \DEG \tilde v_i \geq r(N+1) .\label{eq_sum}
\eeq
Let us now examine $\{v_1 , \ldots , v_{2r}\}$, where $v_1, \ldots
,v_{2r}$ are the rows of $R$. It follows from Lemma~\ref{lem_properties_ring} (ii) that, for $j=1,2$, there are $r$ vectors in $\{v_1 , \ldots , v_{2r}\}$ of leading position $j$ that each have a different order. This implies that there exists a permutation $g$ on $\{1,2,\ldots,2r\}$, such that $\lpos(v_{g(i)}) = \lpos(\tilde
v_i)$ and $\ord(v_{g(i)})= \ord(\tilde v_i)$ for $i=1, \ldots,2r$. Also,
$v_{g(i)}$ can be expressed as a $p$-linear combination of the $\tilde
v_j$'s.
By Theorem~\ref{thm_pGB} the sequence $(\tilde v_1 , \ldots , \tilde v_{2r} )$ has the
$p$-PLM property, so that this linear combination must involve $\tilde v_i$
and it follows that $\lm(v_{g(i)}) \geq \lm(\tilde v_i)$. Since we are
using the {\tt top} monomial ordering, this implies that $\DEG (v_{g(i)})
\geq \DEG(\tilde v_i)$. It now follows from~(\ref{eq_sum}) and
Lemma~\ref{lem_properties_ring} (i) that equality must hold, that is, 
$\DEG (v_{g(i)}) = \DEG (\tilde v_i)$ for $i=1, \ldots , 2r$.
In summary we thus have for $i=1, \ldots , 2r$
\beq
\lm(v_{g(i)}) = \lm(\tilde v_i) \;\;\AND\;\; \ord(v_{g(i)}) = \ord(\tilde v_i) . \label{eq_twoprop}
\eeq
We next prove by induction that $(v_{g(1)} , \ldots , v_{g(2r)})$ is
a $p$-generator sequence whose $p$-span equals $M$.
First ($i=2r$) we observe that we must have $v_{g(2r)}=a_{2r}\tilde v_{2r}$ for
some unit $a_{2r}$. Since $(\tilde v_1 , \ldots , \tilde v_{2r})$ is a
$p$-generator sequence, it follows that
\beq
pv_{g(2r)} = a_{2r}p\tilde v_{2r}=0\label{eq_2rzero}
\eeq
and $\tilde v_{2r} = a_{2r}^{-1}v_{g(2r)} \in \PSPAN \{ v_{g(2r)} \}$. Proceeding
by induction, we assume that for some $i=k+1 \in \{2, \ldots , 2r\}$ the sequence $(v_{g(i)},\cdots, v_{g(2r)})$ is a $p$-generator sequence with
\[
\PSPAN (v_{g(i)},\cdots, v_{g(2r)}) = \PSPAN (\tilde v_i,\cdots, \tilde v_{2r}).
\]
Since $(\tilde v_1 , \ldots ,
\tilde v_{2r})$ is a $p$-basis of $M$, we can write
\[
v_{g(k)} = \sum_{j=1}^{2r} a_j \tilde v_j
\]
for some $a_j \in {\cal A}_p[x]$.
The $p$-PLM property of $(\tilde v_1 , \ldots ,
\tilde v_{2r})$ together with~(\ref{decreasing_degrees_pgrob}), (\ref{decreasing_orders_pgrob}) and~(\ref{eq_twoprop}) implies that $a_j=0$ for $j<k$ and that $a_k$ is a nonzero constant.
%
Thus,
\[
v_{g(k)} = a_k \tilde v_k + v \WITH v\in \PSPAN (\tilde v_{k+1},\cdots,
\tilde v_{2r})\AND a_k \;\mbox{a unit.}
\]
Then $pv_{g(k)} = a_k p\tilde v_k + pv \in \PSPAN(\tilde v_{k+1},\cdots,
\tilde v_{2r})$, so that $pv_{g(k)} \in \PSPAN(v_{g(k+1)},\cdots,
v_{g(2r)})$ by the induction hypothesis. As a result, $\tilde v_k = a_k^{-1} v_{g(k)} - a_k^{-1} v \in \PSPAN \{ v_{g(k)} , \ldots , v_{g(2r)}\}$. In conclusion, for $i=k$ we have $(v_{g(i)} , \ldots , v_{g(2r)})$ is a $p$-generator sequence  and $\PSPAN\{ v_{g(i)} , \ldots , v_{g(2r)}\} = \PSPAN\{ \tilde v_i , \ldots , \tilde v_{2r}\}$ . By induction it now follows that $(v_{g(1)} , \ldots , v_{g(2r)} )$ is a $p$-generator sequence with $\PSPAN\{v_{g(1)} , \ldots , v_{g(2r)}\}=\PSPAN\{\tilde v_1 , \ldots , \tilde v_{2r}\}=M$.
Finally, we prove that $\{v_1 , \ldots , v_{2r}\}$ has the $p$-PLM property. For this, let
\beq
f = a_1 v_1 + \cdots + a_{2r} v_{2r} \label{eq_flincomb}
\eeq
with $a_1, \dots, a_{2r} \in \A_p[x]$. Evidently $\lm (f) \leq \max_{1 \leq i \leq 2r; a_i \neq 0} ( \lm(a_i v_i))$. As a result, in order to prove the $p$-PLM property we need only prove that this upperbound is reached. By grouping together all vectors $a_i v_i$ in~(\ref{eq_flincomb}) that have the same leading position we write
\[
f= f_1 + f_2 ,
\]
where $f_j =0$ if position $j$ is not used in~(\ref{eq_flincomb}). It now follows from the $p$-adic decomposition and (ii) of Lemma \ref{lem_properties_ring} that $\lpos(f_j)=j$ for $j=1,2$ whenever $f_j \neq 0$. More specifically, we then have $\lm(f_j)=\lm(a_{\ell_j}v_{\ell_j})$ for some $\ell_j \in \{1, \ldots , 2r\}$. In case either $f_1=0$ or $f_2=0$ the $p$-PLM property then follows immediately. In case both $f_1$ and $f_2$ are nonzero we recall that their leading positions differ so that, without restrictions, we may assume that $\lm(f_1) < \lm (f_2)$. Then $\lm(f)=\lm(f_2)=\lm(a_{\ell_2}v_{\ell_2})$, which proves the $p$-PLM property.
The property implies, in particular, that $\{v_1, \dots, v_{2r}\}$ is a $p$-basis of $M$.
\pfend
\begin{lem}\label{lem_more_properties_ring}
Let $\Sseq$ be a sequence over $\Zpr$ and let $k \in \{0,\ldots , N\}$. Let $R^k$ be the matrix
obtained by applying Algorithm \ref{alg_ring} to
$S_1, \ldots , S_k$, with rows $v_1^k,\ldots,v_{2r}^k$; denote $v_j^k := \twee{v_{j1}^k}{v_{j2}^k}$ for $j=1,\ldots , 2r$. Then
\begin{itemize}
\item[i)] $v_{j}^k (0) =\twee{0}{0}$ for $j=1,\ldots , r$
\item[ii)] $\ord(v_{j2}^k(0)) =2r-j+1$ for $j=r+1,\ldots , 2r$
\item[iii)] $\lm(v_j^k) \geq \lm(v_{j+1}^k)$ for $j=r+1,\ldots , 2r-1$ with respect to the {\tt top} monomial ordering.
\end{itemize}
\end{lem}
\pf
All conditions are obviously satisfied for $k=0$. Let us now proceed by
induction and assume that the lemma holds for some $k \in \{0,1,\ldots
, N-1\}$.

To prove (i), first note that, by Lemma~\ref{lem_properties_ring} (iii), we have $j \in {\cal P}^k_j$ for $j \in \{1, \dots, r \}$. As a result, for any $j \in \{1, \ldots , r\}$ we have $v^{k+1}_j(x)= x v^k_j(x)$, if $i^*_j=j$, and $v^{k+1}_j(x)=\frac{x}{\theta^{k+1}_{i^*_j}} v^k_{i^*_j}(x)$, otherwise. Thus $v^{k+1}_j(0)=[0 \;\; 0]$.
To prove (ii), let $j \in \{r+1, \dots, 2r \}$. We distinguish two cases:

Case 1: $j \in {\cal P}^k_0$. Then $v^{k+1}_j(x)=v^k_j(x)$ and (ii) follows immediately by induction hypothesis (ii).

Case 2: $j \in {\cal P}^k_{\ell}$ for some $\ell \in \{1, \dots, r\}$, i.e., $\Delta^k_j= \theta^k_j p^{\ell-1}$ for some unit $\theta^k_j$. We distinguish four subcases:

Case 2A: $i^\star_{\ell}= \ell$. Then $\theta^k_{i^\star_{\ell}}=1$ so that $v_j^{k+1}(x)= - \theta^k_j v^k_{\ell}(x) + v^k_j(x) $. Since $v^k_{\ell}(0)=[0 \; \; 0]$ by induction hypothesis (i), it follows that $v^{k+1}_j(0)=v^k_j(0)$ so that (ii) holds by induction hypothesis (ii).

Case 2B: $i^\star_{\ell}= j$. Then again $v_j^{k+1}(x)= - \theta^k_j v^k_{\ell}(x) + v^k_j(x) $ and the reasoning proceeds as in case 2A.

Case 2C: $i^\star_{\ell} > j$. Then $v^{k+1}_j(x)=-\theta^k_j v^k_{i^\star_{\ell}} (x) + \theta^k_{i^*_{\ell}} v^k_j (x)$. By induction hypothesis (ii), $\ord (v^k_{i^\star_{\ell}2}(0)) < \ord(v^k_{j2}(0))$, so that $\ord(v^{k+1}_{j2}(0)) = \ord(v^k_{j2}(0))=2r-j+1$.

Case 2D: $i^\star_{\ell} < j$ and $i^\star_{\ell}\neq \ell$. By definition of $i^\star_{\ell}$ and induction hypothesis (iii) this case cannot happen.

To prove (iii), let $j \in \{r+1, \dots, 2r-1 \}$. Because of Lemma~\ref{lem_pPLM}, we can write $pv^{k+1}_j$ as a $p$-linear combination of $v_1^{k+1}, \ldots , v_{2r}^{k+1}$. Because of (i) and (ii) above, this $p$-linear combination must use $v_{j+1}^{k+1}$ and it follows that $\lm(pv_j^k) \geq \lm(v_{j+1}^k)$ which implies that $\lm(v_j^k) \geq \lm(v_{j+1}^k)$, i.e. (iii) holds.
\pfend
We now present our main result.
\begin{thm}\label{thm_main}
Let $\Sseq$ be a sequence over $\Zpr$ and let $R$ be the matrix
obtained by applying Algorithm \ref{alg_ring} to
$S_1, \ldots , S_N$. Denote the rows of $R$ by $v_1, \ldots
,v_{2r}$; denote $L:= \DEG v_{r+1}$ with respect to the {\tt top} monomial ordering. Then the complexity of the sequence equals $L$ and $v_{(r+1)2}$ is a feedback polynomial of shortest length $L$. More specifically, a parametrization of all shortest length feedback polynomials is given by
\[
a v_{(r+1)2} + \sum_{j \in \{1, \dots, 2r\} \backslash \{r+1\}}a_j v_{j2} ,
\]
with $0\neq a \in \A_p$ and for all $j\neq r+1$ the polynomial $a_j\in \A_p[x]$ chosen such
that $\DEG (a_j) \leq L-\deg v_j$.
Furthermore, let $j^\star$ be such that $\lpos (v_{j^\star})=2$ and $\ord (v_{j^\star})=r$. Let $\tilde L:= \DEG v_{j^\star}$. Then the complexity of the reciprocal sequence $S_N, \dots, S_1$ equals $\tilde L$ 
and $v_{j^\star 2}$ is a minimal characteristic polynomial of $S_N, \dots, S_1$. More specifically, a parametrization of all minimal characteristic polynomials of $S_N, \dots, S_1$ is given by
\beq
a v_{j^\star 2} + \sum_{j \in \{1, \dots, 2r\} \backslash \{j^\star\}}a_j v_{j2} ,\label{eq_para_reciprocal_ring}
\eeq
with $0\neq a \in \A_p$ and for all $j \neq j^\star$ the polynomials $a_j\in \A_p[x]$ chosen such
that $\DEG (a_j) \leq \tilde L - \deg v_j$.
In particular, if $j^\star=r+1$ then $v_{j^\star 2}$ is bidirectional and~(\ref{eq_para_reciprocal_ring}) also parametrizes all bidirectional minimal characteristic polynomials of the reciprocal sequence $S_N, \ldots , S_1$. Otherwise, i.e. if $j^\star \neq r+1$ then any choice of $a_{r+1}\in \A_p [x]$ such that $\DEG (a_{r+1}) \leq \tilde L - \deg v_{r+1}$ and $a_{r+1}(0) \neq 0$ gives a bidirectional minimal characteristic polynomial of $S_N, \ldots , S_1$. 
\end{thm}
\pf
The first parametrization follows immediately from Lemma~\ref{lem_pPLM}, Lemma~\ref{lem_more_properties_ring} and Corollary~\ref{cor_ring}.
Let us now consider the reciprocal sequence in order to prove the second parametrization~(\ref{eq_para_reciprocal_ring}). From Remark \ref{remark} we know that a minimal characteristic polynomial of $S_N, \dots, S_1$ is given by a vector of $M$ with leading position $2$ and order $r$, of minimal degree.
By Lemma~\ref{lem_pPLM} the set $\{v_1, \dots, v_{2r}\}$ is a $p$-basis of $M$ with the $p$-PLM property.
Corollary~\ref{cor_rec_ring} now implies~(\ref{eq_para_reciprocal_ring}).
\pfend
\begin{example}\label{example_reedsS85}
Consider the sequence $6,3,1,5,6$ over the ring $\Z_9$ (as in the example in~\cite{reedsS85}). Application of Algorithm \ref{alg_ring} yields:
\[
\Delta^1=\bmat{c} 1 \\ 3 \\ 6 \\ 0 \emat, \;\; {\cal P}^1_0=\{4\}, \;\; {\cal P}^1_1=\{1\}, \;\; {\cal P}^1_2=\{2,3\},\;\;
 i^*_1=1, \;\; i^*_2=3,
\]
\[
 \;\;\;\;\;\;\;\;\;\; R^1(x)= \bmat{cccc} x & 0 & 0 & 0 \\ 0 & 0 & 5x & 0 \\ 0 & 7 & 1 & 0 \\0 & 0 & 0 & 1 \emat R^0(x)=\bmat{cc} x^2 & 0 \\ 0 & 5x \\ -6x & 1 \\ 0 & 3 \emat;
\]
\[
\Delta^2=\bmat{c} 1 \\ 3 \\ 3 \\ 0 \emat, \;\; {\cal P}^2_0=\{4\}, \;\; {\cal P}^2_1=\{1\}, \;\; {\cal P}^2_2=\{2,3\}, \;\; i^*_1=1, \;\; i^*_2=3,
\]
\[
\;\;\;\;\;\;\;\;\;\; R^2(x)= \bmat{cccc} x & 0 & 0 & 0 \\ 0 & 0 & x & 0 \\ 0 & -1 & 1 & 0 \\0 & 0 & 0 & 1 \emat R^1(x)=\bmat{cc} x^3 & 0 \\ 3x^2 & x \\ -6x & 4x+1 \\ 0 & 3 \emat;
\]
\[
\Delta^3=\bmat{c} 1 \\ 3 \\ 4 \\ 3 \emat, \;\; {\cal P}^3_0=\emptyset, \;\; {\cal P}^3_1=\{1,3\}, \;\; {\cal P}^3_2=\{2,4\}, \;\; i^*_1=3, \;\; i^*_2=4,
\]
\[
\;\;\;\;\;\;\;\;\;\; R^3(x)= \bmat{cccc} 0 & 0 & x/4 & 0 \\ 0 & 0 & 0 & x \\ -4 & 0 & 1 & 0 \\0 & -1 & 0 & 1 \emat R^2(x)=\bmat{cc}3 x^2 & x^2 + 7x \\ 0 & 3x \\ -4x^3-6x & 4x+1 \\ 6x^2 & 8x+3 \emat;
\]
\[
\Delta^4=\bmat{c} 1 \\ 3 \\ 0 \\ 5 \emat, \;\; {\cal P}^4_0=\{3\}, \;\; {\cal P}^4_1=\{1,4\}, \;\; {\cal P}^4_2=\{2\}, \;\; i^*_1=4, \;\; i^*_2=2,
\]
\[
\;\;\;\;\;\;\;\;\;\; R^4(x)= \bmat{cccc} 0 & 0 & 0 & x/5 \\ 0 & x & 0 & 0 \\ 0 & 0 & 1 & 0 \\ -5 & 0 & 0 & 1 \emat R^3(x)=\bmat{cc} 3x^3 & 7 x^2+6x \\ 0 & 3x^2 \\ -4x^3-6x & 4x+1 \\ 0 & 4x^2+3 \emat;
\]
\[
\Delta^5=\bmat{c} 1 \\ 3 \\ 8 \\ 4 \emat, \;\; {\cal P}^5_0=\emptyset, \;\; {\cal P}^5_1=\{1,3,4\}, \;\; {\cal P}^5_2=\{2\}, \;\; i^*_1=4, \;\; i^*_2=2,
\]
\[
\;\;\;\;\;\;\;\;\;\; R^5(x)= \bmat{cccc} 0 & 0 & 0 & x/4 \\ 0 & x & 0 & 0 \\ 0 & 0 & 4 & -8 \\-4 & 0 & 0 & 1 \emat R^4(x)=\bmat{cc} 0 & x^3 + 3x \\ 0 & 3x^3 \\ 2x^3-6x & 4x^2+7x+7 \\ -3x^3 & 3x^2+3x+3 \emat.
\]
By the above theorem, the complexity of the sequence equals $L=3$ and $4x^2+7x+7$ is a shortest length feedback polynomial, normalized to $7x^2+x+1$. 
It is not unique: a parametrization of all normalized shortest length feedback polynomials of length 3 is given by
\beq
7x^2+x+1 + a(x^3 + 3x),\label{eq_para_example}
\eeq
where $a \in \Z_9$. The complexity of the reciprocal sequence $6,5,1,3,6$ equals $\tilde L=3$ and  $x^3+3x$ serves as a minimal characteristic polynomial of $6,5,1,3,6$.
It is not unique, a parametrization of all monic minimal characteristic polynomials of $6,5,1,3,6$ is given by
\[
x^3 + 3x + b(4x^2+7x+7),
\]
where $b \in \Z_9$. For comparison, in our notation, the algorithm of~\cite{reedsS85} produces the matrix
\[
\bmat{cc} \star & \star \\ \star & \star \\ x^3-6x & x^3 + 7x^2+4x+1 \\ -3x^3 & 5x^3 + 3x^2+3 \emat
\]
rather than $R^5(x)$. Thus it produces the shortest feedback polynomial $x^3 + 7x^2+4x+1$. We verify that this polynomial is indeed in our parametrization~(\ref{eq_para_example}), namely for the parameter choice $a=1$.
Note that it follows from the above parametrization~(\ref{eq_para_reciprocal_ring}) that $x^3 + 7x^2+4x+1$ is the unique monic bidirectional minimal characteristic polynomial of $6,5,1,3,6$ that has constant term $1$.
\end{example}

\section{Conclusions}
\label{sec:con}
In his 1969 paper~\cite{massey69} Massey shows that the Berlekamp-Massey algorithm is an efficient algorithm that yields a parametrization of all shortest feedback shift registers for a given finite sequence $S_1, \ldots,S_N$ of numbers in a field. The main contribution of our paper is an iterative algorithm that yields such a parametrization when $S_1, \ldots,S_N$ are numbers in a finite ring $\Zpr$. Although relying on nontrivial theories of $p$-Gr\"obner bases and $p$-linear dependence, the algorithm is highly practical as we illustrated in an example. It is thus shown in this paper that it is possible to have as much "grip" on this fundamental problem in the ring case as in the field case, despite the existence of zero divisors. 
\ruimte
Existing methods for the ring case, such as in~\cite{reedsS85,interlandoPE97,nechaev92,norton99,kurakin98,michalevN96} yield a solution but no parametrization. For the field case (any field, not just $\Z_p$), our algorithm turns out to be a normalized version of the Gr\"obner-based iterative algorithm of~\cite{fitzpatrick95}.
\ruimte
We have shown that our algorithm simultaneously produces all shortest feedback shift registers for the reciprocal sequence $S_N, \ldots,S_1$. This then implies some additional results on bidirectional shortest feedback shift registers. For the field case these results imply findings in~\cite{salagean09}.
\ruimte
We illustrated our result with the $\Z_9$ example $6,3,1,5,6$ from~\cite{reedsS85}. The corresponding module has an "easy" minimal Gr\"obner basis consisting of $2$ elements---a more interesting example is its subsequence $6,3,1$ which has a minimal Gr\"obner basis consisting of $4$ elements. Using Theorem~\ref{thm_main}, we conclude from the matrix $R^3$ in Example~\ref{example_reedsS85} that the sequence $6,3,1$ has highest complexity possible, namely $L=3$, whereas its reciprocal sequence $1,3,6$ has complexity $\tilde L=2$ and minimal characteristic polynomial $x^2+7x$. A parametrization of all monic minimal characteristic polynomials of $1,3,6$ is given by $x^2+7x + b(8x+3)$, where $b\in \Z_9$. This parametrization clearly does not contain any bidirectional characteristic polynomials.
\ruimte
In our view, Gr\"obner bases are ideal tools for these types of problems because they lead to transparent proofs. We emphasize that we only use Gr\"obner bases conceptually, at no stage do we call upon complex computational packages to compute those bases. Instead we iteratively construct minimal Gr\"obner bases in an efficient way. In fact, since our approach combines transparency and efficiency, one may want to conclude that there is little reason for employing noniterative methods such as the euclidean algorithm or noniterative Gr\"obner basis computation.

\section*{Acknowledgment}
The first author would like to thank two institutions for providing support and a stimulating scientific environment during two separate visits in 2011, namely the University of Aveiro, Portugal (June 2011) and the Centre Interfacultaire Bernoulli at EPFL Lausanne, Switzerland (September 2011).

\bibliographystyle{plain}
\bibliography{code}

\end{document}